\documentclass[%
 amsmath,amssymb,
 aps,
]{revtex4-1}

\usepackage{graphicx}
\usepackage{dcolumn}
\usepackage{bm}

\begin{document}

\title{ Black hole as topological insulator (I): the BTZ black hole case}
\author{Jingbo Wang}
\email{ shuijing@mail.bnu.edu.cn}
\affiliation{Institute for Gravitation and Astrophysics, College of Physics and Electronic Engineering, Xinyang Normal University, Xinyang, 464000, P. R. China}
 \date{\today}
\begin{abstract}
Black holes are extraordinary massive objects which can be described classically by general relativity, and topological insulators are new orders of matter that could be use to built a topological quantum computer. They seem to be different objects, but in this paper, we claim that the black hole can be considered as kind of topological insulator. For BTZ black hole in three dimensional $AdS_3$ spacetime we give two evidences to support this claim: the first evidence comes from the  black hole ``membrane paradigm", which says that the horizon of black hole behaves like an electrical conductor. On the other hand, the vacuum can be considered as an insulator. The second evidence comes from the fact that the horizon of BTZ black hole can support two chiral massless scalar field with opposite chirality. Those are two key properties of 2D topological insulator. We also consider the coupling with the electromagnetic field to show that the boundary modes can conduct the electricity. For higher dimensional black hole the first evidence is still valid. So we conjecture that the higher dimensional black hole can also be considered as higher dimensional topological insulators. This conjecture will have far-reaching influences on our understanding of quantum black hole and the nature of gravity.
\end{abstract}
\bibliographystyle{unsrt}
\maketitle
\section{Introduction}
Black holes are massive objects. They have so strong gravitational field that even the light cannot escape. Their existence in universe is supported by many observational evidences, the most direct one comes from the gravitational waves \cite{abb1}. Even through the black holes are solutions of classical general relativity, the quantum effect cannot be neglected near the horizon. Indeed the seminal work of Bekenstein \cite{bek1} and Hawking \cite{hawk1} show that the black holes are actually prefect black body which has temperature and entropy. Up to now, we still don't have a complete quantum theory of black hole.

The black hole ``membrane paradigm" \cite{mb1,mb2} says that to an outside observer, the horizon behaves like a viscous fluid. Besides, it is also electrical conductive. On the other hand, the vacuum inside the horizon can be considered as insulator. This property is similar to the topological insulator.

Topological insulators \cite{ti1,ti2} are new orders of matter. Unlike the usual orders which are associated with broken symmetry, this new order has topological origin, and can be described by topological field theory \cite{tqft1}. Roughly speaking, a topological insulator is a bulk insulator but has conducting boundary states. They exist both in two \cite{2d1,2d2,2d3} and three dimensions \cite{3d1,3d2,3d3}. It was show \cite{tibf1} that BF theory is the effective theory for topological insulators in 2D and 3D just as the Chern-Simons theory for the quantum Hall effect \cite{wen1,wen2}. This BF theory can describe key properties of topological insulators, including the elementary excitation and their statics, the edge theory and so on.

In this paper, we show that the BTZ black hole can be considered as two dimensional topological insulator, or quantum spin Hall state. The paper is organized as follows. In section II, the BTZ black hole in $AdS_3$ is analyzed. We give two evidences to support the claim that BTZ black hole can be considered as topological insulators. In section III, the coupling with the electromagnetic field is considered to show that the boundary modes can indeed conduct the electricity.In section IV, we conjecture that higher dimensional black holes also can be considered as higher dimensional topological insulators. Section V is the conclusion.

\section{BTZ black hole}
In three dimensional spacetime, thing becomes easier and more clear. When there exist a negative cosmological constant $\Lambda<0$, the black hole solution--which is called BTZ black hole \cite{btz1}-- can exist. Due to the membrane paradigm \cite{mbbtz1}, the conductivity of the horizon is
\begin{equation}\label{1}
    \sigma=\frac{1}{g^2},
\end{equation}
where $g$ is the gauge coupling constant. Thus the horizon can be considered as metallic and the vacuum as insulator. This is just the salient property of 2D topological insulator.

Another property of topological insulator is that they have two chiral bosonic modes flowing in opposite direction. Those boundary modes appear because the boundary break the gauge symmetry. Those boundary states are also appeared in gravity theory, which was studied long time ago \cite{edge1,edge2}. In Ref.\cite{whcft1}, it was shown that the horizon of BTZ black hole can support a pair of chiral massless scalar field with opposite chirality which is the same as topological insulator.

In this section we give another derivation for this result which is familiar for condensed matter physicist \cite{wen1,tibf1}. One can get this result in three steps: Firstly, as shown in Ref.\cite{at1,witten1}, $(2+1)-$dimensional general relativity with $\Lambda=-\frac{1}{L^2}$ can be cast into $SO(2,1)\times SO(2,1)$ Chern-Simons theory.
One can define two SO$(2,1)$ connection 1-form
\begin{equation}\label{a1}
    A^{(\pm)a}=\omega^a\pm \frac{1}{L} e^a,
\end{equation}where $e^a$ and $\omega^a$ are the co-triad and spin connection 1-form respectively. Then up to boundary term, the first order action of gravity can be rewritten as
\begin{equation}\label{a2}
    I_{GR}=\frac{1}{8\pi G}\int e^a \wedge ({\rm d}\omega_a+\frac{1}{2}\epsilon_{abc}\omega^b \wedge \omega^c)-\frac{1}{6L^2}\epsilon_{abc}e^a\wedge e^b \wedge e^c\\=I_{CS}[A^{(+)}]-I_{CS}[A^{(-)}],
\end{equation}
where $A^{(\pm)}=A^{(\pm)a}T_a$ are SO$(2,1)$ gauge potential, and the Chern-Simons action is
\begin{equation}\label{a3}
    I_{CS}=\frac{k}{4\pi}\int Tr\{A\wedge {\rm d}A+\frac{2}{3}A\wedge A \wedge A\},
\end{equation}
with
\begin{equation}\label{a4}
    k=\frac{L}{4G}.
\end{equation}
One can get the CS equation
\begin{equation}\label{a5}
    F^\pm={\rm d}A^{(\pm)}+A^{(\pm)} \wedge A^{(\pm)}=0.
\end{equation}

The equation implies that the potential $A$ can be locally written as
\begin{equation}\label{a6}
    A=g^{-1}{\rm d}g.
\end{equation}

Secondly, it is well known that on a manifold with boundary, the Chern-Simons theory reduces to a chiral Wess-Zumino-Novikov-Witten(WZNW) theory on the boundary \cite{witten2}.
On a manifold with the form $M=R\times \Sigma$, the Chern-simons action can be written canonically as
\begin{equation}\label{a12}
    I_{CS}=-\frac{k}{4\pi}\int dx^0 \int_{\Sigma} \varepsilon^{ij} Tr(A_i \dot{A}_j-A_0 F_{ij}).
\end{equation}
Choose the gauge $A_0=0$ gives the constrain
  \begin{equation}\label{a13}
    F_{ij}=0.
\end{equation}
This constrain can be solved by $A_i=g^{-1}\partial_i g$. Submit this solution into the action (\ref{a12}) one can get
\begin{equation}\label{a14}
    I_{CS}=\frac{k}{4\pi}\int_{\partial M} Tr(g^{-1}\partial_0 g g^{-1}\partial_1g)+\frac{k}{12\pi}\int_M Tr(g^{-1}{\rm d}g)^3,
\end{equation}which is just the chiral WZNW action. This is valid for arbitrary boundary.

The final step is to consider the horizon as boundary. To study the field theory on the horizon of BTZ black hole, it is more appropriate to adopt the advanced Eddington coordinate. The metric of BTZ black hole is
\begin{equation}\label{a7}
    ds^2=-N^2 dv^2+2 dv dr+r^2 (d\varphi+N^\varphi dv)^2.
\end{equation}
Choose the following co-triads
\begin{equation}\label{a8}
    l_a=-\frac{1}{2}N^2 dv+dr,\quad n_a=-dv,\quad m_a=r N^\varphi dv+r d\varphi,
\end{equation}
which gives the following connection:
\begin{equation}\label{a9}
    A^{-(\pm)}=-(N^\varphi\mp \frac{1}{L})dr-\frac{N^2}{2} d(\varphi\pm\frac{v}{L}),\quad  A^{+(\pm)}=-d(\varphi\pm\frac{v}{L}),\quad A^{2(\pm)}=r (N^\varphi\pm\frac{1}{L})  d(\varphi\pm\frac{v}{L}).
\end{equation}
Define new coordinates,
\begin{equation}\label{a10}
 u=\varphi-\frac{v}{L}, \quad  \tilde{u}=\varphi+\frac{v}{L}.
\end{equation}
A crucial property of the connection is that, on the whole manifold, one has
\begin{equation}\label{a11}
    A^{(+)}_u\equiv 0,\quad A^{(-)}_{\tilde{u}}\equiv 0.
\end{equation}

Let's consider the right sector $A^{(+)}$. In following we omit the superscript ${(+)}$. Choose $u$ as time coordinate and $i={r,\tilde{u}}$ as the other coordinates. For a general SO$(2,1)$ group element $g(u,i)$, using the Gauss decomposition, it can be written as
\begin{equation}\label{a15}
    g=\left(
                  \begin{array}{cc}
                    1 & \frac{1}{\sqrt{2}}x_1 \\
                    0 & 1 \\
                  \end{array}
                \right)
                \left(
                         \begin{array}{cc}
                           e^{\Psi_1/2} & 0 \\
                           0 & e^{-\Psi_1/2} \\
                         \end{array}
                       \right)
                       \left(
                         \begin{array}{cc}
                           1 & 0 \\
                           -\frac{1}{\sqrt{2}}y_1 & 1 \\
                         \end{array}
                       \right).
\end{equation}
Within this parameter, the WZNW action (\ref{a14}) is
\begin{equation}\label{a16}
    kI_{WZNW}=\frac{k}{4\pi}\int_{\partial M}du d\tilde{u}\frac{1}{2}(\partial_u \Psi \partial_{\tilde{u}} \Psi-e^\Psi (\partial_u x \partial_{\tilde{u}} y+\partial_u y \partial_{\tilde{u}} x)).
  \end{equation}
Let's solve the constraint equation (\ref{a13}) near the horizon. The property of horizon is encode in $N^2=0$. Near the horizon $r=r_+$, a small parameter $\epsilon=r-r_+$ can be defined, and $N^2\approx 2 \kappa \epsilon$, so
\begin{equation}\label{a17}
    A^-_{\tilde{u}}\approx-\kappa \epsilon.
\end{equation} On the other hand,
\begin{equation}\label{a18}
    A^-_{\tilde{u}}=e^{\Psi_1}\partial_{\tilde{u}}x_1/\sqrt{2},
\end{equation}
and $\Psi_1$ is finite at horizon, so one have
\begin{equation}\label{a19}
    x_1\sim \epsilon,
\end{equation}
which vanish on the horizon. So the final action on the horizon is
\begin{equation}\label{a20}
    kI_{WZNW}=\frac{k}{4\pi}\int_{\partial M}du d\tilde{u}\frac{1}{2}\partial_u \Psi_1 \partial_{\tilde{u}} \Psi_1\\
    =\frac{k}{4\pi L}\int_{\partial M}d\varphi dv [(\partial_v \Psi_1)^2- L^2 (\partial_{\varphi} \Psi_1)^2],
  \end{equation}
with $\Psi_1$ depending only on $\tilde{u}=\varphi+\frac{v}{L}$. So it is an action for chiral massless scalar field $\Psi_1$.

The similar results can be get for the left sector $A^{(-)}$, which gives another chiral massless scalar field $\Psi_2$ depending only on $u=\varphi-\frac{v}{L}$. Thus, on the horizon there exist two chiral massless scalar field with opposite chirality, the same as the 2D topological insulator \cite{tibf1}.

Combined with the ``membrane paradigm", we can confirm that BTZ black hole can be considered as 2D topological insulator.
\section{Coupled with electromagnetic field}
To interpret the physical meaning of field $\Psi_1$, we couple the whole system to background electromagnetic(EM) field $a_\mu$. We mainly follow the method in Ref.\cite{tong}. To keep the $U(1)$ gauge symmetry of EM field, the coupling term is chosen to be \cite{fntw}
\begin{equation}\label{a21}
    S_j=\int d^3 x \epsilon^{\mu\nu\rho}A^2_\mu \partial_\nu a_\rho.
\end{equation}

We will set $a_r=0$ and $a_u,a_{\tilde{u}}$ both are independent of $r$ direction. From Eq.(\ref{a15}) it can be shown that $A^2_i=(\frac{\partial \Psi_1}{\partial r}+e^{\Psi_1} y_1 \frac{\partial x_1}{\partial r}, \frac{\partial \Psi_1}{\partial \tilde{u}}+e^{\Psi_1} y_1 \frac{\partial x_1}{\partial \tilde{u}})$, and $A^2_u=0$.
Insert into coupling term Eq.(\ref{a21}) one can get
\begin{equation}\label{a22}\begin{split}
    S_j=\int d^3 x A^2_r (\partial_u a_{\tilde{u}}-\partial_{\tilde{u}} a_u)
    =\int d^3 x  (\frac{\partial \Psi_1}{\partial r}+e^{\Psi_1} y_1 \frac{\partial x_1}{\partial r})(\partial_u a_{\tilde{u}}-\partial_{\tilde{u}} a_u)\\
    =\int_{r=r_+} d^2 x(\Psi_1+e^{\Psi_1} y_1 x_1) (\partial_u a_{\tilde{u}}-\partial_{\tilde{u}} a_u)+\int d^3 x [x_1 \partial_r (e^{\Psi_1} y_1)(\partial_u a_{\tilde{u}}-\partial_{\tilde{u}} a_u)].
\end{split}\end{equation}
Since we want to consider the fields on the boundary, the last bulk term is dropped out. And on the horizon, $x_1=0$, so the left boundary term is
\begin{equation}\label{a23}\begin{split}
    S'_j=\int_{r=r_+} d^2 x \Psi_1 (\partial_u a_{\tilde{u}}-\partial_{\tilde{u}} a_u)=-\int_{r=r_+} d^2 x a_u \partial_{\tilde{u}} \Psi_1,
\end{split}\end{equation}
which we use the part integral and the condition $\partial_u \Psi_1=0$.

Now define new field
\begin{equation}\label{a24}
    \rho(\tilde{u})=\frac{\partial \Psi_1}{\partial \tilde{u}}
\end{equation}
and switch back to origin coordinate $(v,\varphi)$, one can get
\begin{equation}\label{a25}\begin{split}
    S'_j=\frac{1}{2}\int_{r=r_+} d^2 x (a_v L \rho-a_\varphi \rho).
\end{split}\end{equation}
Now the physical meaning of $\rho$ field is clear, that it represent the charge density and current along the boundary.

Compared with the fields in quantum Hall effect \cite{tong}, the $\rho$ field is the same, and the $\Psi_1$ is similar to $\phi$ in quantum Hall effect. But a crucial difference is that $\phi$ is compact but $\Psi_1$ is not, due to the non-compactness of $SO(1,1)$.
\section{Higher dimensional black holes}
In higher dimensional spacetime $(D\geq 4)$, general relativity can't be reformulated as CS theory, so we can't follow the process in the above section. But the membrane paradigm is still valid, thus the horizon can conduct  electrical current. For example, in four dimension, the surface electrical resistivity is
\begin{equation}\label{5}
    \rho=4\pi \alpha \hbar/e^2=2\alpha R_H,
\end{equation}
where $\alpha$ is the finer structure constant and $R_H$ is the Hall resistivity. And the vacuum is also an insulator, so this suggest that higher dimensional black holes can also be considered as topological insulators.

In higher dimension, the general relativity can be reformulated as constraint BF theory \cite{bf1,bf2}. It was shown that there are gauge degrees of freedom on the boundary \cite{edge2}. But it is still unclear what is the effective theory to describe those degrees of freedom. On the other hand, the effective theory for topological insulators is abelian BF theory, which can give the correct boundary theory for topological insulators \cite{tibf1}.
\section{Conclusion}
In this paper, we claim that black holes can be considered as topological insulators. For BTZ black hole in $AdS_3$ spacetime, we give two evidences to support this claim. The first comes from the black hole membrane paradigm, and the second comes form the fact that on the horizon there exist two chiral massless scalar field with opposite chirality. We also consider the coupling with electromagnetic field to show that the boundary modes can conduct the electricity. The first evidence is also valid for higher dimensional black holes. The analogy between black hole horizon and quantum Hall states was investigated in Ref.\cite{vaid1}, but in our opinion, it is more appropriate to consider black hole as topological insulator.

If this claim is true for all kind of black holes, it will have far-reaching influence on our understanding of quantum black hole and the nature of gravity. Since we have well understanding on the topological insulator, including the band structure, the lattice models, the modified Dirac equation description and so on \cite{tibook}. Those properties can be translated to the quantum black hole.
Also since the topological insulator is a new phase of matter, the formation of black hole can be considered as a topological phase transition without symmetry breaking.

General relativity can be formulated as constraint BF theory, and the effective theory of 2D and 3D topological insulators is also BF theory. This suggest that maybe the gravity is also an effective theory of some more fundamental objects. Thus gravity is an ``emergent" phenomenon \cite{emer1,emer2,emer3,emer4}.

\acknowledgments
 This work is supported by the NSFC (Grant No.11647064) and Nanhu Scholars Program for Young Scholars of XYNU.

\bibliography{it1}

\begin{thebibliography}{10}

\bibitem{abb1}
B.~P. Abbott et~al.
\newblock {Observation of gravitational waves from a binary black hole merger}.
\newblock {\em Phys. Rev. Lett.}, 116(6):061102, 2016.

\bibitem{bek1}
J.~D. Bekenstein.
\newblock Black holes and entropy.
\newblock {\em Physical Review D}, 7(8):2333--2346, 1973.

\bibitem{hawk1}
S.~W. Hawking.
\newblock Black-hole explosions.
\newblock {\em Nature}, 248(5443):30--31, 1974.

\bibitem{mb1}
T.~Damour.
\newblock {Black hole eddy currents}.
\newblock {\em Phys. Rev.}, D18:3598--3604, 1978.

\bibitem{mb2}
K.~S. Thorne, R.~H. Price, and D.~A. Macdonald, editors.
\newblock {\em Black Holes: The Membrane Paradigm}.
\newblock Yale University Press, London, 1986.

\bibitem{ti1}
M.~Z. Hasan and C.~L. Kane.
\newblock {Topological Insulators}.
\newblock {\em Rev. Mod. Phys.}, 82:3045, 2010.

\bibitem{ti2}
J.~E. Moore.
\newblock The birth of topological insulators.
\newblock {\em Nature}, 464(194), 2010.

\bibitem{tqft1}
D.~Birmingham, M.~Blau, M.~Rakowski, and G.~Thompson.
\newblock {Topological field theory}.
\newblock {\em Phys. Rept.}, 209:129--340, 1991.

\bibitem{2d1}
C.~L. Kane and E.~J. Mele.
\newblock {Z-2 Topological Order and the Quantum Spin Hall Effect}.
\newblock {\em Phys. Rev. Lett.}, 95:146802, 2005.

\bibitem{2d2}
B.~A. {Bernevig}, T.~L. {Hughes}, and S.-C. {Zhang}.
\newblock {Quantum Spin Hall Effect and Topological Phase Transition in HgTe
  Quantum Wells}.
\newblock {\em Science}, 314:1757, December 2006.

\bibitem{2d3}
M.~{K{\"o}nig}, S.~{Wiedmann}, C.~{Br{\"u}ne}, A.~{Roth}, H.~{Buhmann}, L.~W.
  {Molenkamp}, X.-L. {Qi}, and S.-C. {Zhang}.
\newblock {Quantum Spin Hall Insulator State in HgTe Quantum Wells}.
\newblock {\em Science}, 318:766, November 2007.

\bibitem{3d1}
L.~Fu, C.~Kane, and E.~Mele.
\newblock {Topological Insulators in Three Dimensions}.
\newblock {\em Phys. Rev. Lett.}, 98(10):106803, 2007.

\bibitem{3d2}
J.~E. {Moore} and L.~{Balents}.
\newblock {Topological invariants of time-reversal-invariant band structures}.
\newblock {\em Phys. Rev. B}, 75(12):121306, March 2007.

\bibitem{3d3}
D.~{Hsieh}, D.~{Qian}, L.~{Wray}, Y.~{Xia}, Y.~S. {Hor}, R.~J. {Cava}, and
  M.~Z. {Hasan}.
\newblock {A topological Dirac insulator in a quantum spin Hall phase}.
\newblock {\em Nature}, 452:970--974, April 2008.

\bibitem{tibf1}
G.~Y. Cho and J.~E. Moore.
\newblock {Topological BF field theory description of topological insulators}.
\newblock {\em Annals Phys.}, 326:1515--1535, 2011.

\bibitem{wen1}
X.-G. Wen.
\newblock {Theory of the edge states in fractional quantum Hall effects}.
\newblock {\em Int. J. Mod. Phys.}, B6:1711--1762, 1992.

\bibitem{wen2}
X.-G. {Wen}.
\newblock {Topological orders and edge excitations in fractional quantum Hall
  states}.
\newblock {\em Advances in Physics}, 44:405--473, September 1995.

\bibitem{btz1}
M.~Banados, C.~Teitelboim, and J.~Zanelli.
\newblock {The Black hole in three-dimensional space-time}.
\newblock {\em Phys. Rev. Lett.}, 69:1849--1851, 1992.

\bibitem{mbbtz1}
N.~Iqbal and H.~Liu.
\newblock {Universality of the hydrodynamic limit in AdS/CFT and the membrane
  paradigm}.
\newblock {\em Phys. Rev.}, D79:025023, 2009.

\bibitem{edge1}
A.~P. Balachandran, L.~Chandar, and A.~Momen.
\newblock {Edge states in gravity and black hole physics}.
\newblock {\em Nucl. Phys.}, B461:581--596, 1996.

\bibitem{edge2}
A.~Momen.
\newblock {Edge dynamics for BF theories and gravity}.
\newblock {\em Phys. Lett.}, B394:269--274, 1997.

\bibitem{whcft1}
J.~Wang and C.-G. Huang.
\newblock {The Conformal Field Theory on the Horizon of BTZ Black Hole}.

\bibitem{at1}
A.~Achucarro and P.~K. Townsend.
\newblock {A Chern-Simons Action for Three-Dimensional anti-De Sitter
  Supergravity Theories}.
\newblock {\em Phys. Lett.}, B180:89, 1986.

\bibitem{witten1}
E.~Witten.
\newblock {(2+1)-Dimensional Gravity as an Exactly Soluble System}.
\newblock {\em Nucl. Phys.}, B311:46, 1988.

\bibitem{witten2}
E.~Witten.
\newblock {Quantum Field Theory and the Jones Polynomial}.
\newblock {\em Commun. Math. Phys.}, 121:351--399, 1989.

\bibitem{tong}
D.~Tong.
\newblock {Lectures on the Quantum Hall Effect}.
\newblock 2016.

\bibitem{fntw}
E.~Fradkin, C.~Nayak, A.~Tsvelik, and F.~Wilczek.

\bibitem{bf1}
J.~F. Plebanski.
\newblock {On the separation of Einsteinian substructures}.
\newblock {\em J. Math. Phys.}, 18:2511--2520, 1977.

\bibitem{bf2}
M.~Celada, D.~Gonzlez, and M.~Montesinos.
\newblock {$BF$ gravity}.
\newblock {\em Class. Quant. Grav.}, 33(21):213001, 2016.

\bibitem{vaid1}
D.~Vaid.
\newblock {Quantum Hall Effect and Black Hole Entropy in Loop Quantum Gravity}.
\newblock 2012.

\bibitem{tibook}
S.-Q. Shen.
\newblock {\em Topological Insulators: Dirac Equation in Condensed Matters}.
\newblock Springer, New York, 2012.

\bibitem{emer1}
A.~D. Sakharov.
\newblock {Vacuum quantum fluctuations in curved space and the theory of
  gravitation}.
\newblock {\em Sov. Phys. Dokl.}, 12:1040--1041, 1968.
\newblock [Gen. Rel. Grav.32,365(2000)].

\bibitem{emer2}
T.~Jacobson.
\newblock {Thermodynamics of space-time: The Einstein equation of state}.
\newblock {\em Phys. Rev. Lett.}, 75:1260--1263, 1995.

\bibitem{emer3}
T.~Padmanabhan.
\newblock {Thermodynamical Aspects of Gravity: New insights}.
\newblock {\em Rept. Prog. Phys.}, 73:046901, 2010.

\bibitem{emer4}
E.~P. Verlinde.
\newblock {On the Origin of Gravity and the Laws of Newton}.
\newblock {\em JHEP}, 04:029, 2011.

\end{thebibliography}
\end{document}